\documentclass[conference]{IEEEtran}
\IEEEoverridecommandlockouts
\usepackage{cite}
\usepackage{amsmath,amssymb,amsfonts}
\usepackage{epsfig}
\usepackage{algorithmic}
\usepackage{graphicx}
\usepackage{textcomp}
\usepackage{xcolor}
\def\BibTeX{{\rm B\kern-.05em{\sc i\kern-.025em b}\kern-.08em
    T\kern-.1667em\lower.7ex\hbox{E}\kern-.125emX}}

\usepackage{caption}
\usepackage{subcaption}
\usepackage{graphicx}
\usepackage{enumitem}
\usepackage{xcolor}
\usepackage{amsmath,amssymb}
\usepackage{hyperref}
\hypersetup{
    colorlinks=true,
    linkcolor=blue,
    filecolor=blue,      
    urlcolor=blue,
    pdftitle={Overleaf Example},
    pdfpagemode=FullScreen,
    }
\newcommand\eg{e.g.}

\begin{document}

\title{Flexible Variable-Rate Image Feature Compression for Edge-Cloud Systems}

\author{\IEEEauthorblockN{Md Adnan Faisal Hossain, Zhihao Duan, Yuning Huang, Fengqing Zhu}
\IEEEauthorblockA{\textit{Elmore Family School of Electrical and Computer Engineering} \\
\textit{Purdue University}\\
West Lafayette, IN, 47906, USA \\
\{hossai34, duan90, huan1781, zhu0\}@purdue.edu}
}

\maketitle

\begin{abstract}
Feature compression is a promising direction for coding for machines.
Existing methods have made substantial progress, but they require designing and training separate neural network models to meet different specifications of compression rate, performance accuracy and computational complexity.
In this paper, a flexible variable-rate feature compression method is presented that can operate on a range of rates by introducing a rate control parameter as an input to the neural network model.
By compressing different intermediate features of a pre-trained vision task model, the proposed method can scale the encoding complexity without changing the overall size of the model.
The proposed method is more flexible than existing baselines, at the same time outperforming them in terms of the three-way trade-off between feature compression rate, vision task accuracy, and encoding complexity. 
We have made the source code available at \href{https://github.com/adnan-hossain/var_feat_comp.git}{github.com/adnan-hossain/var-feat-comp}.
\end{abstract}


\section{Introduction}
\label{sec:intro}

The use of Artificial Intelligence (AI) for applications such as the Internet of Things (IoT), visual surveillance, and autonomous drones often calls for a need to deploy mobile or wearable devices that can process data in real-time. With the advent of light-weight machine learning models~\cite{howard2019searching, sandler2018mobilenetv2} and small graphical processing units(GPUs) specifically designed for mobile devices there has been a general trend towards the use of machine learning models on these mobile devices. However, often it is necessary to deploy more powerful deep neural networks to perform complex tasks (\eg, object detection) or to enhance task performance. This raises an issue, as computational and energy constraints on small mobile devices limit their ability to host these powerful models.    



A common approach to solve this dilemma is to transfer data from the edge (mobile or wearable devices) to the cloud (a local server or a data center) and execute the AI models on the cloud instead, forming an \textit{edge-cloud system}.
Typically, for visual data such as images and videos, a pre-designed image/video codec is adopted on the edge device to compress the data before transmission.
However, most existing codecs(either handcrafted~\cite{taubman2002jpeg2000, lainema2012intra} or learned~\cite{balle2016end, agustsson2017soft, balle2018variational, lee2018context}) are generally designed to reconstruct images with good visual quality and hence are not optimized for vision tasks \cite{duan2022efficient}.
This work instead focuses on \textit{feature compression}~\cite{singh2020end, cohen2020lightweight, matsubara2022supervised, datta2022low, duan2022efficient}, which has the main advantage of being able to improve the compression performance for the specific vision task the compression model is trained on.

 
\begin{figure}[t]
\begin{subfigure}[b]{1.0\linewidth}
    \centering
    \includegraphics[width=\linewidth]{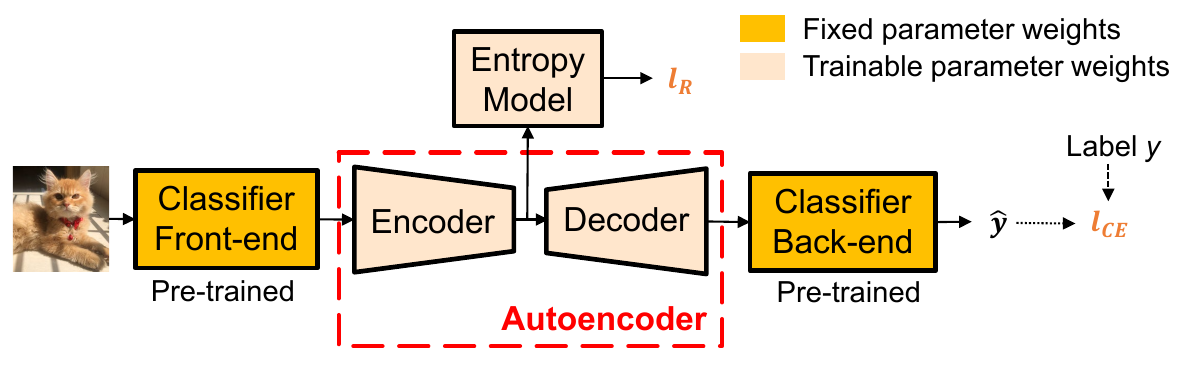}
    \vspace{-0.7cm}
    \caption{Model architecture during training}
\end{subfigure}
\begin{subfigure}[b]{1.0\linewidth}
    \centering
    \includegraphics[width=\linewidth]{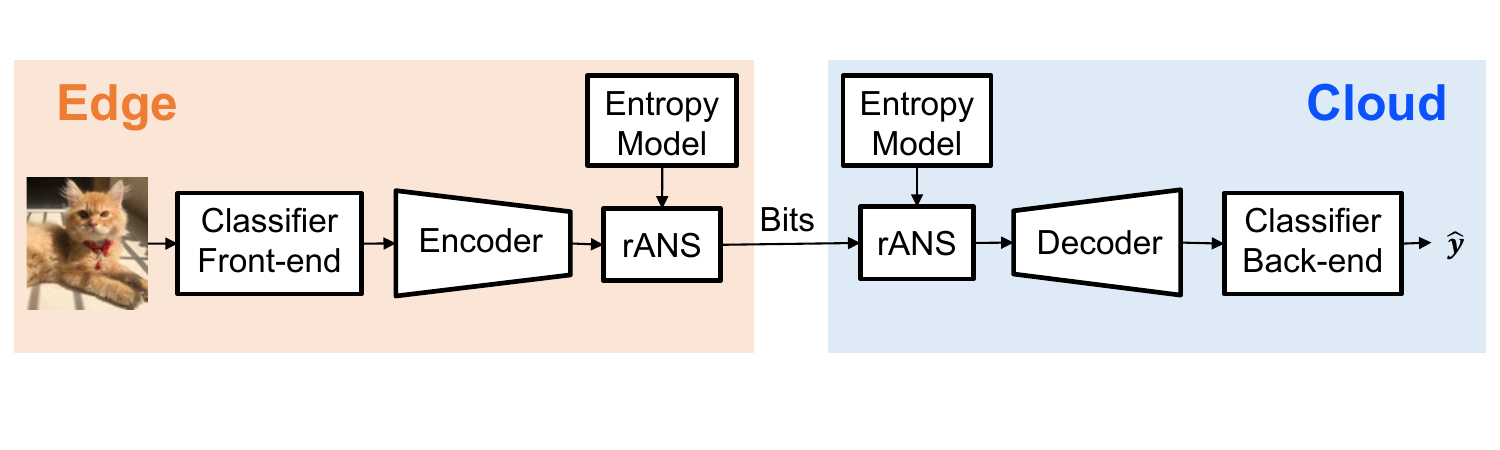}
    \vspace{-1.1cm}
    \caption{Model architecture during deployment}
\end{subfigure}
\caption{
Overview of our method. We compress the intermediate features of a pre-trained vision task model using a variable-rate compressive autoencoder, by which we achieve flexible feature compression for an edge-cloud system.}
\label{fig:model_architecture}
\end{figure}



\begin{figure*}[t]
     \centering
     \begin{subfigure}[t]{0.80\textwidth}
         \centering
         \includegraphics[width=\textwidth]{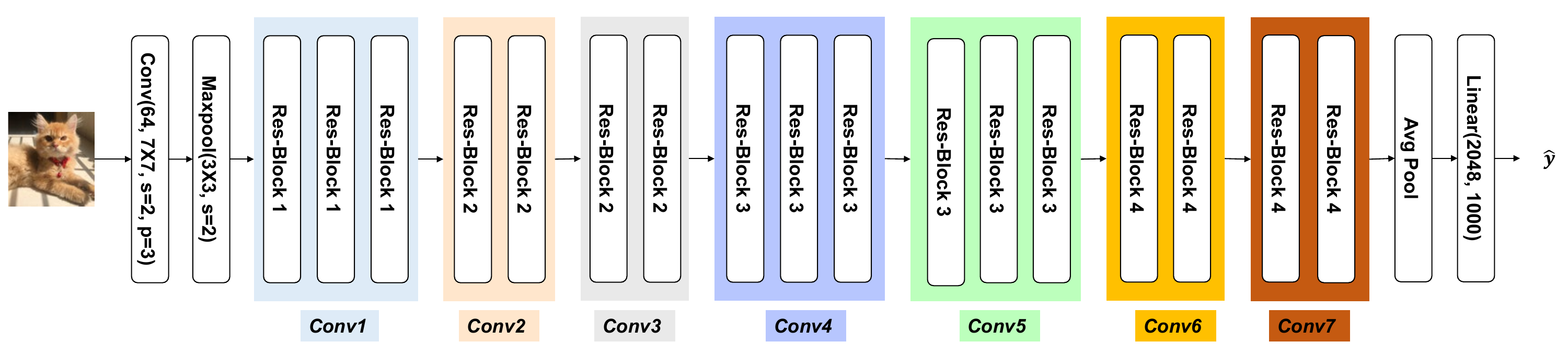}
         \vspace{-0.64cm}
         \caption{Partition of $Convx$ blocks in the ResNet-50 model.}
         \vspace{-0.1cm}
         \label{fig:resnet}
     \end{subfigure}
     \begin{subfigure}[t]{0.80\textwidth}
         \centering
         \includegraphics[width=\textwidth]{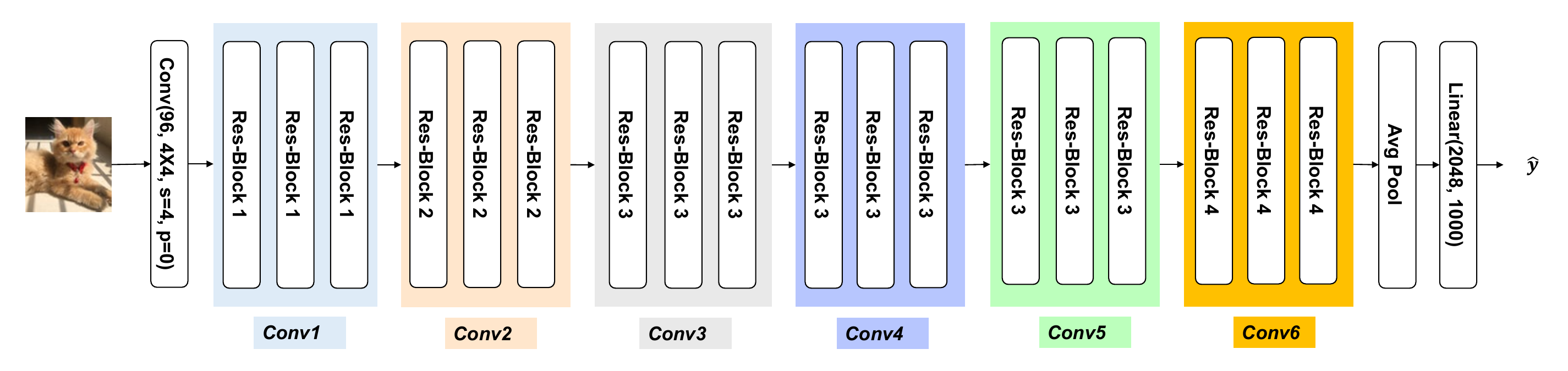}
         \vspace{-0.64cm}
         \caption{Partition of $Convx$ blocks in the ConvNext-T model.}
         \label{fig:convnext}
     \end{subfigure}
\caption{The general framework for partitioning a classifier model into subsequent $Convx$ blocks. The encoding latency is varied by changing the number of $Convx$ blocks at the front end. Note that the network architectures of the Residual Blocks are different for the ResNet-50 and ConvNext-T models.}
\label{fig:split_points}
\end{figure*}

Developing feature compression methods for edge-cloud systems requires consideration of the resource-constrained nature of these systems.
First, the bandwidth-limited channel between the edge and the cloud posits a constraint on the transmission bit rate.
On the other hand, low bit rates are associated with performance degradation for the vision task performed on the cloud.
Finally, due to the variation in computational resources of edge devices, encoders with varying complexity are desired to adapt to different edge devices. Taking these factors into consideration gives rise to a three-way trade-off between rate, accuracy, and complexity (RAC). This trade-off is addressed in \cite{duan2022efficient} by introducing a flexible encoder strategy that can vary in encoding latency by changing the encoder depth. Although the method in \cite{duan2022efficient} can be adapted to different edge devices considering their computational resources, the bandwidth limit is not elegantly addressed since separate models need to be trained for operating on different compression rates.
In this work, we tackle the problem by taking inspiration from the variable-rate image compression model of \cite{choi2019variable} to develop a method to perform variable-rate feature compression in edge-cloud systems.

We present a novel method, overviewed in Fig.~\ref{fig:model_architecture}, to address the RAC trade-off in feature compression for edge-cloud systems.
Our contributions can be summarized as:
\begin{itemize}[noitemsep]
  \item We propose a variable-rate feature compression method for edge-cloud systems. 
  \item We design a dynamic method for distributing a feature compression model between an edge and a cloud device based on specific application scenarios.
  \item We show improved rate-accuracy-complexity trade-off performance compared to previous methods.
\end{itemize}

\section{Preliminaries}
\label{sec:prel}

In this section, we discuss a general framework of learned feature compression for edge cloud systems based on previous works on split computing~\cite{cohen2020lightweight, cohen2021lightweight} and learned feature compression~\cite{matsubara2022supervised, datta2022low, duan2022efficient}. As shown in Fig.~\ref{fig:model_architecture}, a network that has already been trained to perform a certain vision task is split up into a front-end that resides on the edge and a back-end that resides on the cloud using an autoencoder. An entropy model learns to compress the latent feature generated by the encoder network of the autoencoder while the autoencoder is simultaneously trained on a machine vision task, \eg, image classification in our case. The overall training loss is adopted from \cite{singh2020end} and is described as follows:
\begin{equation}
L = l_{CE}(\hat{y}, y) + \lambda l_{R}(\hat{z}) \label{eqn:0}
\end{equation}
where $\hat{y}$ is the output prediction, $y$ is the true label and $l_{CE}(\hat{y}, y)$ is the cross-entropy loss for image classification. In general, this loss can be replaced by the objective loss function for other machine vision tasks. $z$ is the output of the encoder which is also referred to as the bottleneck or latent feature, and $\hat{z}$ is its quantized version. The second component of the loss function is the data-rate or entropy loss $l_{R}(\hat{z})$. $\lambda$ is the Lagrangian multiplier that controls the trade-off between rate and accuracy \cite{singh2020end}.

We choose the factorized entropy model from \cite{balle2016end} as our entropy model and use it to estimate the entropy loss $l_{R}(\hat{z})$. As quantization is a non-differentiable process, during training $\hat{z}$ is estimated by sampling from a uniform distribution of width $1$ centered around the value of $z$. $l_{R}(\hat{z})$ is defined as:

\begin{equation}
l_{R}(\hat{z}) = \mathbb{E} \left[ -\log_{2}p(\hat{z}; \phi) \right] \label{eqn:0_1}
\end{equation}
where $\phi$ represents the parameters of the entropy model, and the expectation is w.r.t. the marginal distribution of $\hat{z}$, which is determined by the training dataset and the encoder.

The goal of the model optimization process is to minimize the loss $L$ to obtain the optimum $\theta^{*}$ and $\phi^{*}$, where $\theta$ denotes the parameters of the autoencoder. This is formulated as:
\begin{equation}
\theta^{*}, \phi^{*} = \operatorname*{arg\,min}_{\theta, \phi} \sum_{x,y\in D}^{} l_{CE}(\hat{y}, y) + \lambda l_{R}(\hat{z}) \label{eqn:0_2} 
\end{equation}
Here, $x$ and $y$ are the input and labels sampled from the training set $D$. For image compression, $x$ consists of training images, but for feature compression $x$ consists of features generated by a learned feature generator such as a neural network. $x$ is passed on to the encoder where it is used to generate the latent feature $z$ according to \eqref{eqn:1}. The entropy model uses $z$ to create its quantized version $\hat{z}$ and the prior probability distribution $p(\hat{z})$. $\hat{z}$ is passed to the decoder to generate the reconstructed feature $\hat{x}$ according to \eqref{eqn:2}.
\begin{eqnarray}
z &=& g_{z}(x;\theta_{enc}) \label{eqn:1} \\
\hat{x} &=& f_{\hat{x}}(\hat{z};\theta_{dec}) \label{eqn:2} 
\end{eqnarray}

During inference, the entropy model is shared between the edge and cloud device. On the edge, the entropy model uses the distribution $p(\hat{z})$ to encode the latent feature into bits using the range based asymmetric numeral system (rANS) algorithm \cite{duda2013asymmetric} and then transmits the bits to the cloud. On the cloud, the entropy model transforms the bits back to $\hat{x}$.

\section{Method}


In order to develop a flexible method for distributing the feature compression model between the edge and the cloud, we first analyze the constraints that govern the decision of choosing the split point (the point in the network where subsequent layers are sent to the cloud and preceding layers are kept in the edge) in a typical neural network-based image classifier, \eg, ResNet-50 \cite{he2016deep} in Sec.~\ref{sec:method_1}. Based on these constraints, a general strategy for splitting up an image classification model between the front-end and the back-end as shown in Fig.~\ref{fig:model_architecture} is described in Sec.~\ref{sec:method_2}. Finally, we describe the design of our autoencoder in Sec.~\ref{sec:method_3} and explain our variable-rate feature compression method in Sec.~\ref{sec:method_4} 

\subsection{Constraints on Choosing the Split Point}
\label{sec:method_1}
Choosing the split point depends on factors such as transmission bandwidth, machine vision task performance and encoding latency. Limited transmission bandwidth requires that a split point be chosen such that the transmitted features are coded with the minimum possible bit rate. However, improving performance accuracy often comes at the cost of coding the data using higher bit rates to reduce information loss. In general for a given depth of the front-end architecture, this trade-off can be tuned by varying the value of the parameter $\lambda$. At the same time, for a fixed $\lambda$ and accuracy, increasing the depth of the front-end architecture should improve bit rate performance as deeper layers are more compressible \cite{choi2022scalable} and contain more usable information~\cite{xu2020theory}. On the other hand, in many real-time application scenarios, it is often desirable to have minimum encoding latency on the edge device which can be achieved by selecting a shallower split point.

\subsection{Strategy for Splitting the Network}
\label{sec:method_2}
Based on the discussion in Sec.~\ref{sec:method_1}, we propose a general strategy for splitting the classification network. As most modern classification networks consist of groups of similar residual blocks cascaded together~\cite{he2016deep, liu2022convnet, szegedy2016rethinking}, the proposed strategy can be generalized. The process is described as follows:
\begin{enumerate}[noitemsep]
  \item The classification network is split into groups of similar residual layers \cite{he2016deep}. These groups of similar residual layers are referred to as \textit{bottleneck blocks}. 
  \item The residual layers of each \textit{bottleneck block} are split into groups of two or three depending on the total number of layers in the block. Each of these groups is called a \textit{Convx} block.
  \item The number of \textit{Convx} blocks in the front-end can be varied. More \textit{Convx} blocks leads to better rate-accuracy performance due to a deeper front-end architecture while fewer \textit{Convx} blocks reduce the encoding latency.
\end{enumerate}

In Fig.~\ref{fig:resnet}, we apply our strategy to split up the ResNet-50 model from \cite{he2016deep} which we use to show the RAC performance of our proposed method in Sec.~\ref{sec:exp}. We also apply our strategy to split up the ConvNext-Tiny model from \cite{liu2022convnet} in Fig.~\ref{fig:convnext} to demonstrate the generalizability of our method. 

\begin{figure}[t]
 \centerline{\epsfig{figure=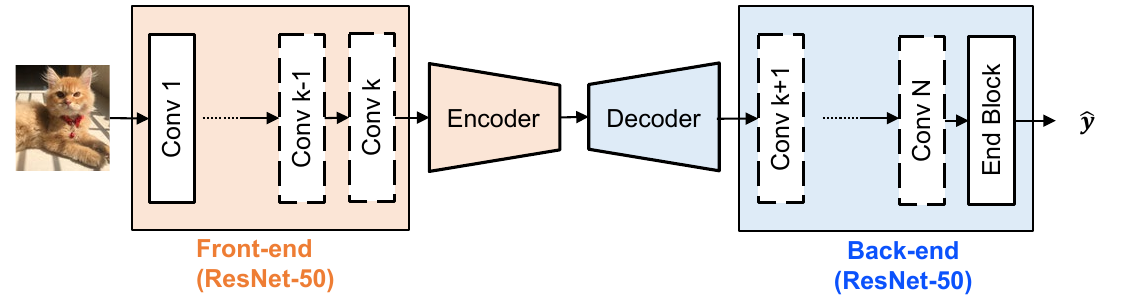, width=\columnwidth}}
\caption{The proposed model configuration Config.$k$, where $k$ is the number of $Convx$ blocks in the encoder. $N$ is the total number of $Convx$ blocks the classifier can be divided into using the strategy proposed in \ref{sec:method_2}. For the ResNet-50 model, $N = 7$. The solid boxes represent layers that are fixed and the dashed boxes represent layers that can be moved between the encoder/decoder for different model configurations.}
\label{fig:modelx}
\end{figure}

In order to achieve a rate-accuracy-complexity (RAC) trade-off, $\lambda$ is varied to tune between bit-rate and classification accuracy, and the encoding complexity is varied by changing the number of \textit{Convx} blocks in the front-end. In the latter part of this discussion, three model configurations are tested and each configuration is referred to as Config.$k$ where $k$ is the number of \textit{Convx} blocks in the front-end. The general model configuration is shown in Fig.~\ref{fig:modelx}.

\begin{figure*}[t]
     \centering
     \begin{subfigure}[t]{0.20\textwidth}
         \centering
         \includegraphics[width=\textwidth]{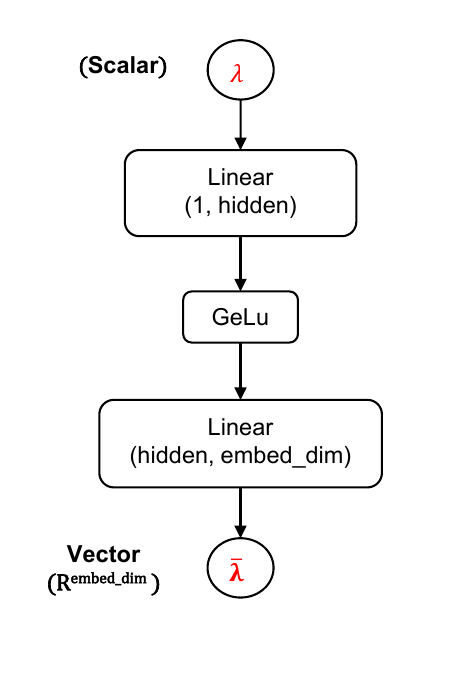}
         \caption{The feed-forward neural network used to generate vector embeddings from $\lambda$}
         \label{fig:embedding}
     \end{subfigure}
     \hfill
     \begin{subfigure}[t]{0.78\textwidth}
         \centering
         \includegraphics[width=\textwidth]{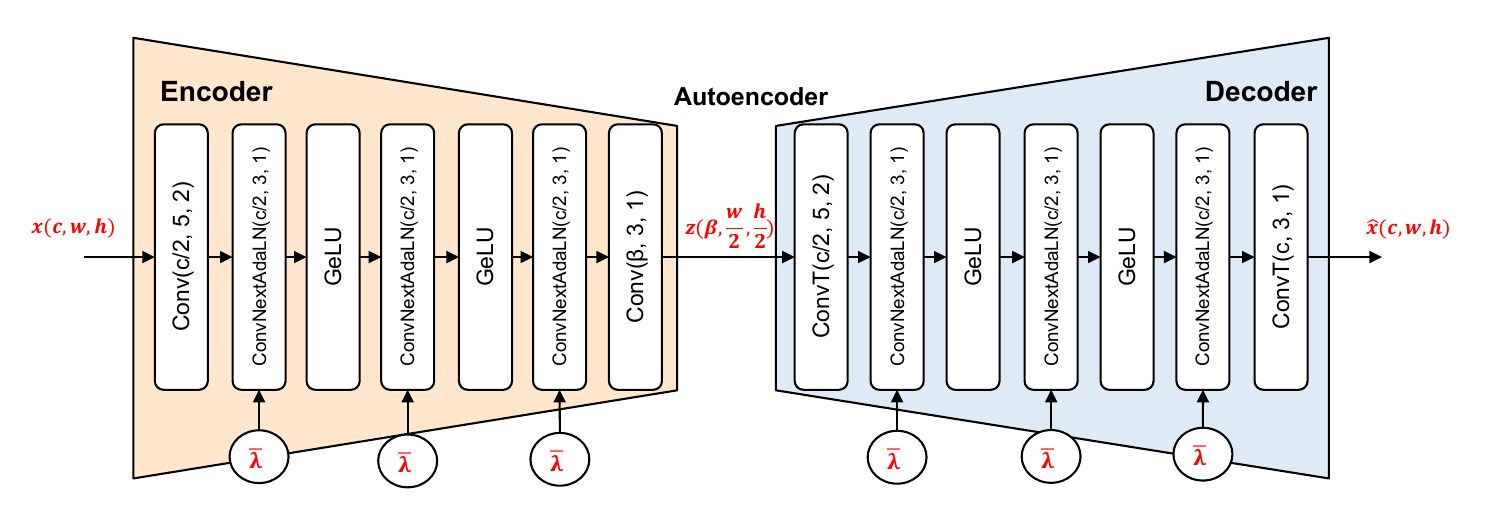}
         \caption{Autoencoder architecture used for variable-rate feature compression. Conv($N$, $k$, $s$) represents a convolution layer with $N$ kernels of size ($k$, $k$) and a stride of $s$ while ConvT($N$, $k$, $s$) represents its transposed convolution. GeLU is the Gelu activation function. ConvNextAdaLN($N$, $k$, $s$) is the network shown in Fig.~\ref{fig:AdaLN}.}
         \label{fig:autoencoder}
     \end{subfigure}
\caption{(a) and (b) illustrate the network architectures used to perform varaible-rate feature compression. $\lambda$ is passed as the input to the network in (a) and its output $\bar{\lambda}$ is used as the input to the network in (b) that is responsible for tuning the bit-rate of the model. The number of filters $\beta$ in the last convolution layer of the autoencoder is varied between model configurations so that the product $\beta * w/2 * h/2$ is always a constant.}
    \label{fig:encoder-decoder}
\end{figure*}

\subsection{Design of Autoencoder}
\label{sec:method_3}
An autoencoder is inserted at the split point to generate latent features at the encoder which are compressed and transmitted to the decoder. At the decoder, the received feature is decompressed and used to reconstruct the original input feature. The autoencoder is designed in a manner so that the same design principles can be applied to all the model configurations. The encoder and decoder have symmetrical structures with an initial convolution layer for downsampling or upsampling, respectively. The last layer of both the encoder and decoder are also convolution layers, but they are used for channel reduction or expansion only. In between the two convolution layers there are three ConvNext Adaptive Layer Normalization (ConvNextAdaLN) blocks adopted from \cite{duan2023qarv}. 

The encoding latency primarily depends on two factors: the depth of the edge model architecture that controls the transmission time of the input image data through the network, and the total number of elements in the bottleneck feature $z$ that dictates the time for the entropy model to perform feature compression. In order to analyze the change in latency due to the depth of the edge network only,  the encoder is designed so that the total number of elements in the bottleneck feature $z$ is constant for all model configurations. This is achieved by tuning the number of kernels in the last convolution layer of the encoder as shown in Fig.~\ref{fig:encoder-decoder}.


\subsection{Variable Rate Feature Compression}
\label{sec:method_4}
In this work, we propose a novel \textit{variable rate feature compression} model. By employing a variable-rate scheme, the proposed model only has to train one single dynamic network which can be deployed in a range of bit rate settings by tuning the rate control parameter $\lambda$.

\begin{figure}[t]
 \centerline{\epsfig{figure=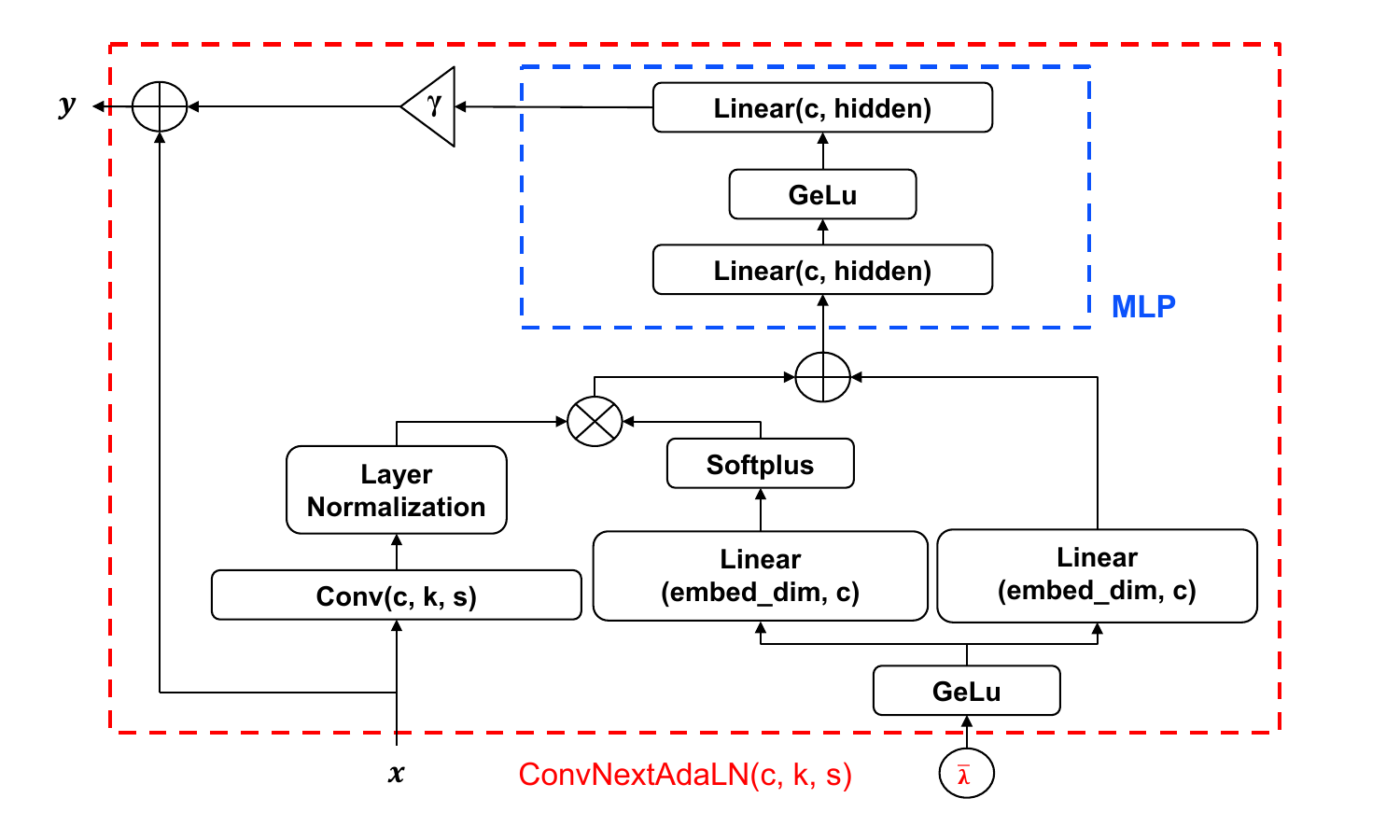, width=\columnwidth}}
\caption{ConvNext Adaptive Layer Normalization block. Conv($N$, $k$, $s$) represents a convolution layer with $N$ kernels, a kernel of size ($k$, $k$) and stride of $s$.}
\label{fig:AdaLN}
\end{figure}

In fixed rate compression models, $\lambda$ is a hyperparameter that is added to the loss function and remains fixed throughout training. However, to operate at variable bit rates, our model incorporates $\lambda$ as an input to the network rather than a hyperparameter. This is achieved by using ConvNextAdaLN blocks from \cite{duan2023qarv} in place of regular convolution layers in the autoencoder network. The network architecture of the ConvNextAdaLN block is shown in Fig.~\ref{fig:AdaLN}. 
The block serves as a conditional convolution layer conditioned on $\lambda$ \cite{choi2019variable, van2016conditional}. Vector embeddings of $\lambda$, referred to as
$\bar{\lambda} \in \mathbb{R}^\text{embed\_dim}$
are passed as the input to the ConvNextAdaLN block along with the output from the previous layer. $\bar{\lambda}$ is generated from $\lambda$ using a feedforward neural network as shown in Fig.~\ref{fig:embedding}. Vector embeddings are commonly used to enhance the performance of neural networks by creating low-dimensional representations of the input, which allows the network to learn the relationships between inputs more efficiently \cite{vaswani2017attention, tellez2019neural,duan2023qarv}.

As $\lambda$ is now an input to the model that is varied during training within the interval $(0, \lambda_\text{max}]$ (referred to as $\Lambda$), the training loss function is modified to \eqref{eqn:3} and the goal of the model is to find the optimum $\theta$ and $\phi$ according to \eqref{eqn:4}: 
\begin{equation}
L = \sum_{\lambda \in \Lambda }^{}l_{CE}(\hat{y}, y; \lambda) + \lambda l_{R}(\hat{z}; \lambda) \label{eqn:3}\\
\end{equation}
\begin{equation}
\theta^{*}, \phi^{*} = \operatorname*{arg\,min}_{\theta, \phi} \sum_{x,y\in D}^{} \sum_{\lambda \in \Lambda}^{}l_{CE}(\hat{y}, y;\lambda) + \lambda l_{R}(\hat{z};\lambda) \label{eqn:4}
\end{equation}


To achieve optimum performance, instead of sampling $\lambda$ from a uniform distribution in the range $(0, \lambda_\text{max}]$ where $\lambda_\text{max} >> 0.01$, $\lambda$ is sampled from a log distribution in the same range. This is done to train the model evenly over high bit-rate and low bit-rate regions as the bit-rate is more evenly distributed in the log space over the range $(0, \lambda_\text{max}]$.

\section{Experiments}
\label{sec:exp}

Our method is compared to two previous methods for feature compression in an edge-cloud setting: the Entropic Student method \cite{matsubara2022supervised} and the method by Duan and Zhu, 2022 \cite{duan2022efficient}. Note that neither of these methods tackle variable-rate feature compression, which is addressed in our method.

\subsection{Experiment Settings}
We use the 1000-class Image Classification dataset of ImageNet (ILSVRC 2012) \cite{deng2009imagenet} containing $1.28$ million training and $50,000$ validation images for training and evaluating our method. All the images are resized to $224\times224$. We train each model for $60$ epochs with a stochastic gradient descent optimizer and an initial learning rate of $0.01$. A cosine learning rate decay is also applied. To enable a fair comparison with the previous methods, we use the ResNet-50 model \cite{he2016deep} as the classification network in all our experiments.  

We measure classification accuracy using the Top-1 accuracy, bit rate as the bits per pixel (bpp), and encoding complexity in terms of GPU encoding latency. In order to visualize a rate-accuracy-complexity trade-off, we use the Delta-accuracy metric introduced in \cite{duan2022efficient}. The Delta-accuracy of a compression method is a measure of its average accuracy improvement (following the convention of BD-rate and BD-PSNR \cite{bjontegaard2001calculation}) over all bit rates with respect to a baseline method.


\subsection{RAC Comparison between Proposed Model Configurations}
We design three network architectures with varying encoder complexity to perform a quantitative comparison between our proposed model configurations: Config.1, Config.2 and Config.3. The general architecture for the model configurations is shown in Fig.~\ref{fig:modelx}. The Delta-accuracy (with respect to Config.1) and encoding latency of the three model configurations are shown in Table \ref{tab:exp1}. From the table, it can be seen that as the depth of the edge network is increased (moving from Config.1 to Config.3), the Delta-accuracy of the model also increases. This is expected, as deeper layers of a classification network have more condensed features which contain more useful information relevant to the classification task \cite{xu2020theory}. Hence, it requires fewer bits to compress features from deeper layers. However, this increase in rate-accuracy performance comes at the cost of higher encoding latency. 

In Table \ref{tab:exp1}, classifier latency is defined as the time in milliseconds required for data to pass through the classifier front-end, compression time refers to the time needed by the entropy model to compress the latent feature, and encoding latency is the total time required for the data to pass through the edge device and get compressed into bits. It can be seen that the compression time is roughly the same between all model configurations due to the design principles adopted for the autoencoder as explained in Sec.~\ref{sec:method_3}. We can also see from the table that the change in encoding latency between configurations (+0.78, +0.59) ms is almost equal to the change in classifier latency (+0.64, +0.59) ms. Therefore, the change in encoding latency between the model configurations is primarily due to the proposed tuning of the number of $Conv x$ blocks in the edge network. By adopting this scheme for varying complexity we can ensure that the latency always increases from shallower configurations to deeper ones.

\begin{figure}[t]
 \centering
\includegraphics[width=0.85\linewidth]{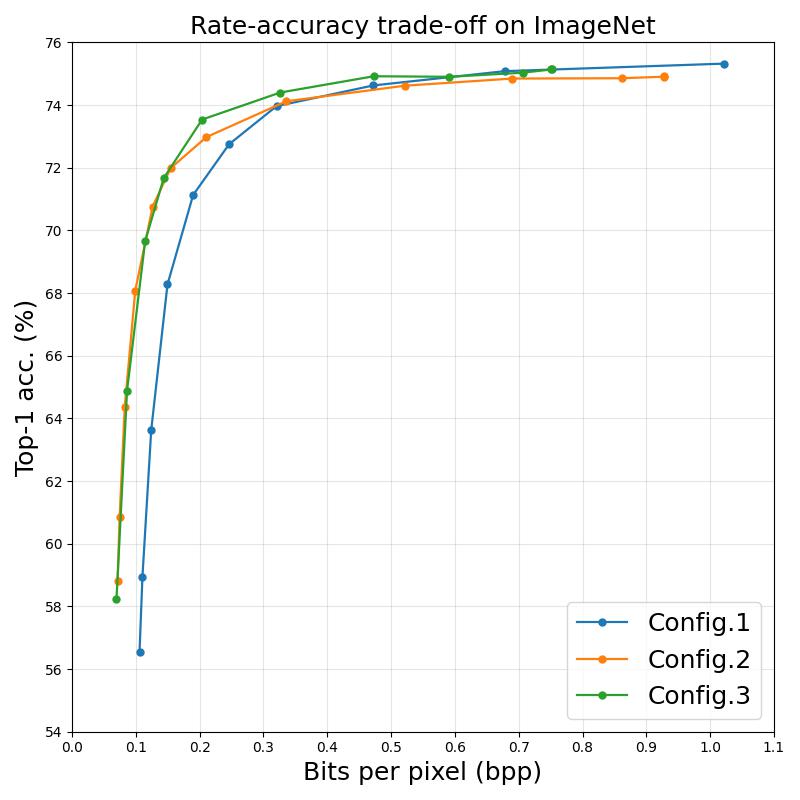}
\vspace{-0.1cm}
\caption{Rate-Accuracy comparison of three different proposed model configurations}
\label{fig:RAC1}
\end{figure}

Fig.~\ref{fig:RAC1} shows the rate-accuracy curves of the three model configurations. The expected trend between the networks is seen mostly in the low and medium bit rate regions [0.05, 0.4] bpp. During the training of these model configurations, $\lambda$ is sampled from a log-uniform distribution. It is important to note that the final rate-accuracy curve relies on the range and technique used for sampling $\lambda$. In this work, the optimal sampling range is empirically found to be [0.0001, 5.12].

\begin{table}[t]
\begin{center}
\caption{Rate-Accuracy-Complexity comparison of three proposed model configurations} \label{tab:exp1}
\vspace{-0.1cm}
\begin{tabular}{c|c|c|c}
  \hline
Configuration &  Config.1 & Config.2 & Config.3
  \\
  \hline
Delta-Acc. (\%) & 0.0 & 0.46 & 1.50 \\         
Classifier Latency (ms) & 0.92  & 1.56 & 2.15 \\      
Compression Time (ms) & 2.17  & 2.25 & 2.25 \\      
Encoding Latency (ms) & 4.42  & 5.20 & 5.79
\end{tabular}
\vspace{-0.3cm}
\end{center}
\end{table}

\subsection{RAC Comparison of Proposed Model Configurations with other Baselines}
\begin{figure}[t]
  \centering
\includegraphics[width=0.85\linewidth]{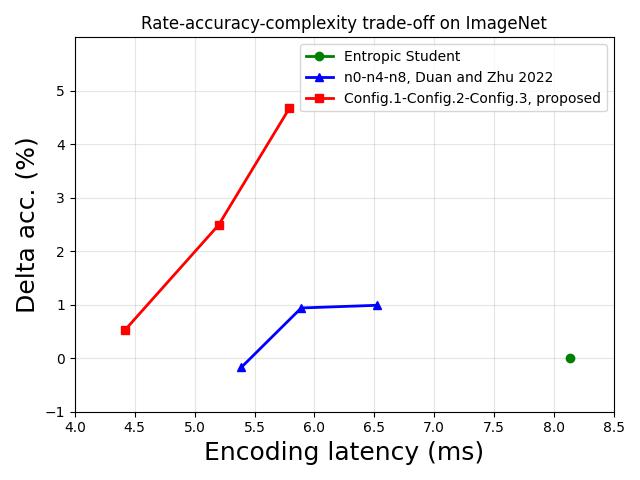}
\vspace{-0.1cm}
\caption{Rate-Accuracy-Complexity performance of our proposed model against other baselines measured in terms of Delta-accuracy(with respect to the Entropic Student model) and encoding latency}
\vspace{-0.3cm}
\label{fig:latency}
\end{figure}

In this section, we compare the proposed method with the Entropic Student method \cite{matsubara2022supervised} and the method of Duan and Zhu, 2022 \cite{duan2022efficient}. The Delta-accuracy value for each method with respect to the Entropic Student method is computed using their individual rate-accuracy curves. The plot of Delta-accuracy vs. Encoding latency for each of the outlined methods is shown in Fig.~\ref{fig:latency}. The encoding complexity of the proposed model increases from Config.1 to Config.2 to Config.3. In Duan and Zhu, 2022, the encoding complexity is varied by changing the number of residual blocks $N$ in their variable encoder architecture. The three architectures used in this study correspond to N = $[0, 4, 8]$ and are referred to as n0, n4 and n8. The Entropic Student model does not include a method to change its complexity. Thus, it only operates in one latency setting.

We can see from Fig.~\ref{fig:latency} that Config.1 has a better rate-accuracy performance (higher Delta-accuracy) compared to n0 (+0.70 \%) and the Entropic Student model (+0.53 \%). It also operates at a lower encoding latency than all the other models. However, it has a lower Delta-accuracy than n4 (-0.41 \%) and n8 (-0.46 \%). On the other hand, both Config.2 and Config.3 outperform all the other models in terms of Delta-accuracy although at the cost of a higher encoding complexity. Config.2 has a Delta-accuracy of 2.49 \% compared to the Entropic Student model and Config.3 has a Delta-accuracy of 4.67 \%.

\section{Conclusion}

In this paper, we propose a learned variable rate feature compression method for edge-cloud systems. Due to the variable rate nature of the method, it is superior to the existing feature compression methods which require different networks to be trained for every rate-accuracy pair. This work also proposes a method for changing the encoder complexity while keeping the overall size of the system roughly constant. We perform experiments to demonstrate the ability of our proposed method to adapt to different rate-accuracy-complexity requirements depending on the application scenario.
We also compare our proposed method with existing baseline methods, and show that our approach outperforms them in terms of RAC trade-off.
A drawback of the proposed method is that a new model has to be trained to operate at each encoder complexity. It is left as future work to develop a single dynamic model that can operate at a range of encoder complexities.

\bibliographystyle{IEEEbib}
{\footnotesize \bibliography{icme2023_main}}

\end{document}